\documentclass[11pt,fleqn,twoside]{article}
\usepackage{amsfonts,amssymb,latexsym}
\makeatletter
\newcommand{\prava}{\footnotesize\it
\begin{flushright}
\begin{minipage}{18cm}
Copyright \copyright 1998 by T.A. Ivanova
\end{minipage}
\end{flushright}}

\newcommand{\name}[1]{\begin{flushleft}
                       \LARGE \bf #1
                       \end{flushleft}\vspace{-3mm}}

\newcommand{\Author}[1]{\begin{flushleft}
                       \it #1 \end{flushleft}}

\newcommand{\Adress}[1]{\begin{flushleft}
                       \it #1 \end{flushleft}}

\newcommand{\Date}[1]{\begin{flushleft}
                      \small  \it #1 \end{flushleft}}

\newcommand{\ehkol}{Author \ name}
\newcommand{\ohkol}{Article \ name}
\renewcommand{\@evenhead}{
\hspace*{-3pt}\raisebox{-15pt}[\headheight][0pt]{\vbox{\hbox to \textwidth
{\thepage \hfil \ehkol}\vskip4pt \hrule}}}
\renewcommand{\@oddhead}{
\hspace*{-3pt}\raisebox{-15pt}[\headheight][0pt]{\vbox{\hbox to \textwidth
{\ohkol \hfil \thepage}\vskip4pt\hrule}}}
\renewcommand{\@evenfoot}{}
\renewcommand{\@oddfoot}{}

\newcommand{\be}{\begin{equation}}
\newcommand{\ee}{\end{equation}}
\newcommand{\ba}{\hspace*{-5pt}\begin{array}}
\newcommand{\ea}{\end{array}}
\newcommand{\p}{\partial}
\newcommand{\ds}{\displaystyle}
\makeatother

\begin{document}
\setcounter{page}{396}
\thispagestyle{empty}

\renewcommand{\ehkol}{T.A. Ivanova}
\renewcommand{\ohkol}{On Inf\/initesimal Symmetries
of the Self-Dual Yang-Mills Equations}

\begin{flushleft}
\footnotesize \sf
Journal of Nonlinear Mathematical Physics \qquad 1998, V.5, N~4,
\pageref{ivanova-fp}--\pageref{ivanova-lp}.
\hfill {\sc Article}
\end{flushleft}

\vspace{-5mm}

\renewcommand{\footnoterule}{}
{\renewcommand{\thefootnote}{} \footnote{\prava}

\name{On Inf\/initesimal Symmetries  of \\
the Self-Dual Yang-Mills Equations}\label{ivanova-fp}

\Author{T.A. IVANOVA}

\Adress{Bogoliubov Laboratory of Theoretical Physics, JINR, \\
141980 Dubna, Moscow Region, Russia\\ E-mail:
ita@thsun1.jinr.dubna.su}

\Date{Received March 18, 1998; Accepted June 4, 1998}

\begin{abstract}
\noindent
Inf\/inite-dimensional algebra of all inf\/initesimal transformations of
solutions of the self-dual Yang-Mills equations is described. It
contains as subalgebras the inf\/inite-dimensional algebras of hidden
symmetries related to gauge and conformal transformations.
\end{abstract}

\section{Introduction}

Yang-Mills theory is a non-Abelian generalization of the Maxwell
theory of electromagnetism.  The dynamics of the non-Abelian gauge
f\/ields is described by the Yang-Mills~(YM) equations, and the study
of the space of solutions to the YM equations is of particular
interest. In 1975 the equations giving a very important subclass of
solutions to the~YM equations were introduced~\cite{ivanova:BPST}. These
equations were called the self-dual Yang-Mills~(SDYM) equations;
their solutions provide absolute minima for the Yang-Mills functional
in Euclidean 4-space.  There exists a large literature on the
geometric meaning of the SDYM equations (see
e.g.~\cite{ivanova:AHS,ivanova:PenR,ivanova:WW,ivanova:MW}).

Our aim is to investigate inf\/initesimal symmetries of the SDYM
equations.  Under a symmetry we understand a transformation which
maps solutions of the SDYM equations into solutions of these
equations. In other words, symmetry transformations preserve the
solution space. It is known that all local symmetries of the SDYM
equations, which are also called manifest symmetries, are given by
gauge transformations and conformal transformations. Since 1979, in a
number of papers ~\cite{ivanova:P-D}, it was shown that the SDYM equations
have nonlocal, so-called `hidden' symmetries which are related to
global gauge transformations. More general gauge-type symmetries were
described in~\cite{ivanova:UN,ivanova:Tak,ivanova:Crane,ivanova:ita}.  In~\cite{ivanova:PP}, an af\/f\/ine
extension of conformal symmetries was introduced. The twistor
interpretation of this algebra was discussed in ~\cite{ivanova:ita}. But the
problem of describing all possible (local and nonlocal) symmetries is
not yet solved.

The paper is organized as follows: In \S\S\,2,3 we recall the main
def\/initions (for more details see e.g.~\cite{ivanova:DV,ivanova:DNF}) and the
Penrose-Ward twistor
correspondence~\cite{ivanova:W,ivanova:AW,ivanova:PenR,ivanova:WW,ivanova:MW}.
In \S\S\,4,5 we give a cohomological description of
the inf\/initesimal symmetries of the SDYM equations by reducing this
problem to the problem of describing inf\/initesimal symmetries of
holomorphic bundles over a twistor space.

\section{Def\/initions and notation}

{\bf 2.1. The SDYM equations.} Let us consider a principal f\/ibre
bundle $P=P({\mathbb R}^4, SU(n))$ over the Euclidean space ${\mathbb
R}^4$ with the structure group $SU(n)$. Let $A_\mu (x)$ be components
of the connection 1-form $A=A_\mu (x)dx^\mu$ in the bundle $P$,
$x\in{\mathbb R}^4$, $\mu ,
\nu ,\ldots =1,\ldots,4$. Here and in what follows summation over
repeated indices is understood.  By def\/inition, components
$F_{\mu\nu}(x)$ of the curvature 2-form
$F=\frac{1}{2}F_{\mu\nu}(x)dx^\mu\wedge dx^\nu$ in $P$ are
$F_{\mu\nu}(x)=[D_\mu , D_\nu ]= [\p_\mu +A_\mu , \p_\nu +A_\nu
]=\p_\mu A_\nu - \p_\nu A_\mu + [A_\mu , A_\nu ],\ \p_\mu :=\p /\p
x^\mu$. Fields $A_\mu$ and $F_{\mu\nu}$ take values in the Lie
algebra $su(n)$.

The {\it self-dual Yang-Mills} (SDYM) {\it equations} have the form
\be
F_{\mu\nu}=\frac{1}{2} \varepsilon _{\mu\nu\rho\sigma}F_{\rho\sigma},
\ee
where $\varepsilon _{\mu\nu\rho\sigma}$ is the completely
antisymmetric tensor on ${\mathbb R}^{4}$ and $\varepsilon
_{1234}=1$.

\medskip

\noindent
{\bf 2.2. Manifest symmetries of the SDYM equations.} As is known,
eqs.~(1) are invariant under the algebra of {\it infinitesimal gauge
transformations}
\be
A_\mu \mapsto A^\vartheta_\mu = A_\mu + \delta_\vartheta A_\mu +
\cdots , \qquad \delta_\vartheta A_\mu =\p_\mu \vartheta + [A_\mu ,
\vartheta ], \qquad \vartheta (x)\in su(n)
\ee
and under the algebra of {\it infinitesimal conformal
transformations}
\be
A_\mu \mapsto A^N_\mu = A_\mu + \delta_N A_\mu + \ldots,\qquad
\delta_N A_\mu =N^\nu\p_\nu A_\mu + A_\nu\p_\mu N^\nu,
\ee
where a vector f\/ield $N=N^\nu\p_\nu$ is any generator
\[
\ba{l}
\ds X_a=\delta_{ab}\eta_{\mu\nu}^{b}x_\mu\partial_\nu ,\qquad
Y_a=\delta_{ab}\bar\eta_{\mu\nu}^{b}x_\mu\partial_\nu ,\qquad P_\mu
=\partial_\mu ,\\[4mm]
\ds K_\mu =\frac{1}{2}x_\sigma x_\sigma \partial_\mu -x_\mu B, \qquad
B=x_\sigma\partial_\sigma,\qquad a,b,\ldots =1,2,3,
\ea
\]
of the 15-parameter conformal group that is locally isomorphic to the
group $SO(5,1)$~\cite{ivanova:DNF}.  Here $\{X_a\}$ and $\{Y_a\}$ generate
two commuting $SO(3)$ subgroups in $SO(4)$, $P_\mu$ are the
translation generators, $K_\mu$ are the generators of special
conformal transformations and~$B$ is the dilatation generator;
$\eta_{\mu\nu}^{a}=
\{\varepsilon_{bc}^{a}, \mu =b, \nu =c;\ \delta_{\mu}^{a}, \nu =4;
\ -\delta_{\nu}^{a}, \mu =4\}$ and $\bar\eta_{\mu\nu}^{a}=
\{\varepsilon_{bc}^{a}, \mu =b, \nu =c;\ -\delta_{\mu}^{a}, \nu =4;
\ \delta_{\nu}^{a}, \mu =4\}$ are the 't\,Hooft tensors satisfying
\[
\frac{1}{2} \varepsilon
_{\mu\nu\rho\sigma}\eta_{\rho\sigma}^{a}=\eta_{\mu\nu}^{a},\qquad
\frac{1}{2} \varepsilon
_{\mu\nu\rho\sigma}\bar\eta_{\rho\sigma}^{a}=-\bar\eta_{\mu\nu}^{a},
\]
i.e. $\eta_{\mu\nu}^{a}$ are the self-dual tensors and
$\bar\eta_{\mu\nu}^{a}$ are the anti-self-dual tensors.

\medskip

\noindent
{\bf 2.3. Complex structure on ${\mathbb R}^4$.} The most general
constant {\it complex structure} $J=(J_{\mu}^{\nu})$ on ${\mathbb
R}^{4}$ has the form
\be
J_{\mu}^{\nu}= s_a\bar\eta_{\mu\sigma}^{a}\delta^{\sigma\nu}\
\Longrightarrow\ J_{\mu}^{\sigma}J_{\sigma}^{\nu}=-\delta_{\mu}^{\nu},
\ee
where real numbers $s_a$ parametrize a two-sphere $S^2$, $s_as_a=1$,
$\bar\eta_{\mu\sigma}^{a}$ are the anti-self-dual 't~Hooft tensor.
Using $J$, one can introduce $(0,1)$ vector f\/ields $\bar V_{\bar 1}$,
$\bar V_{\bar 2}$
\be
\bar V_{\bar 1}=\p_{\bar y}-\lambda\p_z,\qquad \bar V_{\bar
2}=\p_{\bar z}+\lambda\p_y,
\ee
satisfying $J^\nu_\mu \bar V^\mu_{\bar 1,\bar 2} =-i
\bar V^\nu_{\bar 1,\bar 2}$. Here
$y=x_1+ix_2,\ z=x_3-ix_4,\ \bar y=x_1-ix_2,\ \bar z=x_3+ix_4$ are the
complex coordinates on ${\mathbb R}^4\simeq {\mathbb C}^2$, and
$\lambda = (s_1+is_2)/ (1+s_3)$ is a local complex coordinate on
$S^2\simeq {\mathbb C}\!P^1$.

\medskip

\noindent
{\bf 2.4. Twistor space for ${\mathbb R}^4$.} Let $C_+:=\{\lambda\in
{\mathbb C}\!P^1: |\lambda |\le 1+\alpha\}$, where $0<\alpha <1$ is a
positive real number, $C_-:=\{\lambda\in {\mathbb C}\!P^1: |\lambda
|\ge 1-\alpha\}$ (including $\lambda =\infty$).  Then $C_+$ and $C_-$
form a two-set open cover of the Riemann sphere ${\mathbb C}\!P^1$
with the intersection $C_\alpha = C_+\cap C_- = \{\lambda :\ 1-\alpha
\le |\lambda |\le 1+\alpha\}$. The vector f\/ield $\p_{\bar\lambda}:= \p
/\p_{\bar\lambda}$ is antiholomorphic $(0,1)$ vector f\/ield with
respect to the standard complex structure $\varepsilon =
i\;d\lambda\otimes \p_\lambda -i\;d{\bar\lambda}\otimes
\p_{\bar\lambda}$ on ${\mathbb C}\!P^1$.

{\em Twistor space} ${\mathcal Z}$ of ${\mathbb R}^{4}$ is the bundle
$\pi : {\mathcal Z}\to {\mathbb R}^{4}$ of complex structures on
${\mathbb R}^{4}$ associated with the principal $SO(4)$-bundle of
orthogonal frames of ${\mathbb R}^{4}$~\cite{ivanova:AHS}.  It means that the
f\/ibre $\pi^{-1}(x)$ of ${\mathcal Z}\to {\mathbb R}^{4}$ over a point
$x\in{\mathbb R}^{4}$ coincides with the space ${\mathbb C}\!P^1$ of
complex structures on ${\mathbb R}^{4}$ def\/ined above.  The space
${{\mathcal Z}}$ is the trivial bundle over ${\mathbb R}^{4}$ with
f\/ibre ${\mathbb C}\!P^1$, hence ${{\mathcal Z}}= {\mathbb
R}^{4}\times {\mathbb C}\!P^1$ is a manifold which can be covered by
two coordinate patches:
\setcounter{equation}{5}
\renewcommand{\theequation}{\arabic{equation}{\rm a}}
\be
{{\mathcal Z}}= U_+\cup U_-,\qquad
U_+:=\{x\in {\mathbb R}^{4}, \lambda\in C_+\},\qquad U_-:=\{x\in
{\mathbb R}^{4}, \lambda\in C_-\}
\ee
with the intersection
\setcounter{equation}{5}
\renewcommand{\theequation}{\arabic{equation}{\rm b}}
\be
 U:= U_+\cap U_-=\{x\in {\mathbb R}^4, \lambda\in C_\alpha =C_+\cap
C_-\}.
\ee
Let us denote the cover (6) by ${\mathfrak U}$.

The twistor space ${\mathcal Z}$ is a complex manifold with the
complex structure ${\mathcal J}=(J,\varepsilon )$ on~${\mathcal Z}$.
Vector f\/ields $\bar V_{\bar 1}$, $\bar V_{\bar 2}$ from (5) and $\bar
V_{\bar 3}=
\p_{\bar\lambda}$ are the vector f\/ields on ${\mathcal Z}$ of type
$(0,1)$ with respect to the complex structure ${\mathcal J}$.

\section{The Penrose-Ward correspondence}

{\bf 3.1. Complex vector bundle $\tilde E$ over the twistor space.}
Let $E=P\times_{SU(n)}{\mathbb C}^n$ be a complex vector bundle
associated to $P$. Sections of this bundle are ${\mathbb C}^n$-valued
functions depending on $x\in {\mathbb R}^{4}$.

By using the projection $\pi :\ {\cal Z}
\rightarrow {\mathbb R}^4$, we can pull back the bundle $E$ with the
connection $D=dx^\mu D_\mu $ to the bundle $\tilde E:=\pi^*E$ over
${\mathcal Z}={\mathbb R}^4\times {\mathbb C}\!P^1$. By def\/inition of
the pull back, the pulled back connection $\tilde{D}:=\pi^*D$ will be
f\/lat along the f\/ibres ${\mathbb C}\!P_{x}^{1}$ of the bundle
${{\mathcal Z}}\rightarrow {\mathbb R}^{4}$ and, therefore, the
components of $\tilde{A}:=\pi^*A$ along the vector f\/ields
$\p_\lambda$, $\p_{\bar\lambda}$ in ${\mathbb C}\!P_{x}^{1}$ can be
set equal to zero.  Then we have $\tilde{D}= D+d\lambda\p_\lambda+
d\bar\lambda\p_{\bar\lambda}$.

Local sections of the complex vector bundle $\tilde E$ are ${\mathbb
C}^n$-valued functions def\/ined on open subsets of ${\mathcal
Z}={\mathbb R}^4\times {\mathbb C}\!P^1$.

\medskip

\noindent
{\bf 3.2. Linear system for the SDYM equations and holomorphic
bundles.} Let $\tilde{D}_{\bar a}^{(0,1)}\ (a=1,2,3)$ be components
of $\tilde{D}$ along the $(0,1)$ vector f\/ields $\bar V_{\bar a}$ on
${\mathcal Z}$. A section $\xi$ of the bundle $\tilde{E}$ is called a
{\it local holomorphic section} if it is a local solution of the
equations $\tilde{D}_{\bar a}^{(0,1)}\xi =0$ or, in local coordinates
on ${\mathcal Z}$,
\setcounter{equation}{6}
\renewcommand{\theequation}{\arabic{equation}{\rm a}}
\be
 (D_{\bar y} -\lambda D_z)\xi (x,\lambda,\bar\lambda )=0,
\ee
\setcounter{equation}{6}
\renewcommand{\theequation}{\arabic{equation}{\rm b}}
\be
(D_{\bar z} +\lambda D_y)\xi (x,\lambda ,\bar\lambda )=0,
\ee
\setcounter{equation}{7}
\renewcommand{\theequation}{\arabic{equation}}
\be
\p_{\bar\lambda}\xi (x,\lambda ,\bar\lambda )=0.
\ee
The equations $\tilde{D}_{\bar a}^{(0,1)}\xi =0$ on sections $\xi$
of the complex vector bundle $\tilde{E}$ def\/ine a {\em holomorphic
structure} in $\tilde{E}$. Accordingly, the bundle $\tilde E$ is said
to be {\em holomorphic} if eqs. (7), (8) are compatible, i.e. the (0,2)
components of the curvature of the bundle $\tilde E$ are equal to
zero.

The solution of eq. (8) is $\xi (x,\lambda )$.  Equations (7) on $\xi
(x,\lambda )$ are called the {\it linear system} for the SDYM
equations~\cite{ivanova:BZ,ivanova:W}.  It is easy to see that the compatibility
conditions of the linear system (7) coincide with the SDYM equations
written in the complex coordinates $y,z,\bar y,\bar z$ on ${\mathbb
R}^4\simeq{\mathbb C}^2$.

Equations (7) have local solutions $\xi_\pm(x,\lambda )$ over
$U_\pm\subset{\mathcal Z}$, and $\xi_+=\xi_-$ on $U=U_+\cap U_-$ (for
def\/initions of $U_\pm$ and $U$ see (6)). We can always represent
$\xi_\pm$ in the form $\xi_\pm =\psi_\pm\chi_\pm$, where $\psi_\pm$
are matrices of fundamental solutions of (7) def\/ining a
trivialization of $\tilde E$ over $U_\pm$, and $\chi_\pm\in {\mathbb
C}^n$ are \v{C}ech f\/ibre coordinates satisfying $\bar V_{\bar
a}\chi_\pm =0$ and $ \chi_-={\mathcal F}\chi_+ $ on $U=U_+\cap
U_-\subset {\mathcal Z}$.  The matrix ${\mathcal
F}=\psi_-^{-1}\psi_+$ is the transition matrix in the bundle $\tilde
E$, i.e. holomorphic $SL(n,C)$-valued function on $U$ with
non-vanishing determinant satisfying the conditions on transition
matrices~\cite{ivanova:GH}.

\medskip

\noindent
{\bf 3.3. Ward's theorem.} So, starting from the complex vector
bundle $E$ over ${\mathbb R}^4$ with the self-dual connection $D$, we
can construct the holomorphic vector bundle $\tilde E$ over
${\mathcal Z}$ with the transition matrix ${\mathcal
F}=\psi_-^{-1}\psi_+$ def\/ined on $U\subset{\mathcal Z}$.

Conversely, if we are given the holomorphic vector bundle $\tilde
E=\tilde P({{\mathcal Z}}, SL(n, {\mathbb C} ))\times_{SL(N, {\mathbb
C})}{\mathbb C}^n$ associated to the principal f\/ibre bundle $\tilde
P$ over ${\mathcal Z}$, which is holomorphically trivial on each
f\/ibre ${\mathbb C}\!P_{x}^{1}: \ \tilde E\!\mid_{{\mathbb
C}\!P_{x}^{1}}\simeq {\mathbb C}\!P_{x}^{1}\times {\mathbb C}^n$
(Ward's twistor construction~\cite{ivanova:W}), then on ${\mathbb
C}\!P_{x}^{1}$ the transition matrix ${\mathcal F}$ can be factorized
in the form (Birkhof\/f's theorem):
\[
 {\mathcal F}=\psi^{-1}_-(x,\lambda )\psi_+(x,\lambda ),
\]
 where $\psi_\pm (x,\lambda )$ are $SL(n, {\mathbb C})$-valued
functions holomorphic in $\lambda^{\pm 1}$ on $C_\pm$.

{}From the holomorphicity of ${\mathcal F}$ on $U$ ($\bar V_{\bar
a}{\mathcal F}=0$) it follows that $(\bar V_{\bar
a}\psi_+)\psi_+^{-1}=(\bar V_{\bar a}
\psi_-)\psi_-^{-1}$ and, therefore,
\setcounter{equation}{9}
\renewcommand{\theequation}{\arabic{equation}{\rm a}}
\be
 (\p_{\bar y}\psi_+ - \lambda \p_z\psi_+)\psi_+^{-1}= (\p_{\bar
y}\psi_- - \lambda \p_z\psi_-)\psi_-^{-1}= -(A_{\bar y}(x) - \lambda
A_z(x)),
\ee
\setcounter{equation}{9}
\renewcommand{\theequation}{\arabic{equation}{\rm b}}
\be
 (\p_{\bar z}\psi_+ + \lambda \p_y\psi_+)\psi_+^{-1}= (\p_{\bar
z}\psi_- + \lambda \p_y\psi_-)\psi_-^{-1}= -(A_{\bar z}(x) + \lambda
A_y(x)),
\ee
 and the potentials $\{A_\mu\}$ def\/ined by (10) satisfy the SDYM
equations and do not change after transformations:
$\psi_\pm\mapsto\psi_\pm h_\pm$, where $h_\pm$ are regular
holomorphic matrix-valued functions on $U_\pm$.  This means that the
bundles with transition matrices $h_{-}^{-1}{\mathcal F}h_+$ and
${\mathcal F}$ are holomorphically equivalent.

Let us summarize the facts about the Penrose-Ward correspondence in
the theorem:
\medskip

\noindent
 {\bf Theorem~\cite{ivanova:AW,ivanova:AHS}.} {\it There is a
one-to-one correspondence between gauge equivalence classes of
solutions to the SDYM equations in the Euclidean 4-space and
equivalence classes of holomorphic vector bundles $\tilde E$ over the
twistor space ${\mathcal Z}$, which are holomorphically trivial over
each real projective line ${\mathbb C}\!P_{x}^{1}$ in ${\mathcal
Z}$.}

\setcounter{equation}{10}
\renewcommand{\theequation}{\arabic{equation}}

\section{Inf\/initesimal gauge-type symmetries}

{\bf 4.1. The algebras $C^0({{\mathfrak U}}, {\mathcal H})$ and
$C^1({{\mathfrak U}}, {\mathcal H})$.} We consider the principal
f\/ibre bundle $\tilde P= \tilde P ({{\mathcal Z}}, SL(N,{\mathbb C}))$
over the twistor space ${\mathcal Z}$ and the associated bundle
$Ad\tilde P=\tilde P
\times_{Ad\,SL(n,{\mathbb C})}sl(n,{\mathbb C})$ with the adjoint action of the group
$SL(n,{\mathbb C})$ on the algebra $sl(n,{\mathbb C})$: $\xi\mapsto
Ad_g\xi=g\xi g^{-1}$, $g\in SL(n,{\mathbb C})$, $\xi\in sl(n,{\mathbb
C})$. Let ${\mathcal H}$ be a sheaf of germs of holomorphic sections
of the bundle $Ad\tilde P$ (for def\/inition see ~\cite{ivanova:GH}), $\
\Gamma({\mathcal U},{\mathcal H})$ be a set of all sections of the sheaf
${\mathcal H}$ over an open set ${\mathcal U}\subset {\mathcal Z}$.

A collection $\{\varphi_+, \varphi_-\}$ of sections of ${\mathcal H}$
over the open sets $U_+$ and $U_-$ from (6a) is called a {\it
0-cochain over} ${\mathcal Z}$, subordinate to the cover ${\mathfrak
U}=\{U_+,U_-\}$. Thus, a 0-cochain is an element of the space
\[
C^0({{\mathfrak U}}, {\mathcal H}):= \Gamma (U_+, {\mathcal H})
\oplus \Gamma (U_-, {\mathcal H}).
\]

The space of {\it 1-cochains} with values in ${\mathcal H}$
\[
C^1({{\mathfrak U}}, {\mathcal H}):= \Gamma (U, {\mathcal H})
\]
is a set of sections $\varphi$ of the sheaf ${\mathcal H}$ over
$U=U_+\cap U_-$.  Notice that $C^0({{\mathfrak U}}, {\mathcal H})$ and
$C^1({{\mathfrak U}}, {\mathcal H})$ are Lie algebras of holomorphic
maps: $U_\pm \to sl(n, {\mathbb C})$ and $U
\to sl(n, {\mathbb C})$ respectively with pointwise commutator.

\medskip

\noindent
{\bf 4.2. Action of $C^1({{\mathfrak U}}, {\mathcal H})$ on
transition matrices.} The standard action of the algebra
$C^0({{\mathfrak U}}, {\mathcal H})$ on the space of holomorphic
transition matrices ${\mathcal F}$:
\[
 \delta{\mathcal
F}=\varphi_-{\mathcal F}- {\mathcal F}\varphi_+
\]
gives us holomorphically equivalent bundles.  Hence, these
transformations are trivial.  But we shall consider the action of the
algebra $C^1({{\mathfrak U}}, {\mathcal H})$ on ${\mathcal F}$:
\be
\delta_\varphi{\mathcal F}=
\varphi (\lambda ){\mathcal F}+ {\mathcal
F}\varphi^\dagger\left(-\frac{1}{\bar\lambda }\right),
\ee
where $\varphi\in C^1({{\mathfrak U}},{\mathcal H})$,
$\varphi=\varphi (\lambda )\equiv \varphi (y-\lambda\bar z,
z+\lambda\bar y, \lambda )$, $\varphi (-1/\bar\lambda )\equiv \varphi
(y+\bar z/\bar\lambda , z-\bar y/\bar\lambda , -1/\bar\lambda )$, and
$\dagger$ denotes Hermitian conjugation.

Transformations (11) preserve the holomorphicity of ${\mathcal F}$
and preserve the hermiticity of the bundle $E$; they are local
inf\/initesimal transformations of the transition matrix.

\medskip

\noindent
{\bf 4.3. Inf\/initesimal gauge-type transformations of self-dual
connections.} Let us introduce the $sl(n, {\mathbb C})$-valued smooth
function $\phi$ on $U$:
\[
\phi :=\psi_-(\delta_\varphi{\mathcal F})\psi_{+}^{-1}=
\psi_-\varphi (\lambda )\psi_{-}^{-1}+ \psi_+\varphi^\dagger
\left(-\frac{1}{\bar\lambda}\right)\psi_{+}^{-1},
\]
which is holomorphic in
$\lambda\in C_\alpha$ and can be expanded in Laurent series
\[
\ba{l}
\ds \phi =\sum\limits_{n=-\infty}^{\infty}\lambda^n\phi_n(x)= \phi_-
- \phi_+ ,\\[4mm]
\ds
\phi_+:=\tilde\phi_{0}(x)-\!\sum\limits_{n=1}^{\infty}\lambda^n\phi_n(x),
\quad \!\!
\phi_-:=\hat\phi_{0}(x)+\!\!\sum\limits_{n=-\infty}^{-1}
\lambda^n\phi_n(x),\quad\!\!
\ds \hat\phi_{0}(x)-\tilde\phi_{0}(x)=\phi_{0}(x).
\ea
\]
The splitting $\phi = \phi_- - \phi_+$ is a solution of the
inf\/initesimal variant of the Riemann-Hilbert problem, and functions
$\phi_\pm\in sl(n, {\mathbb C})$ are holomorphic in $\lambda\in
C_\pm$. It follows from (10) that $\tilde D^{(0,1)}_{\bar a}\phi =0$,
and, therefore,
\setcounter{equation}{11}
\renewcommand{\theequation}{\arabic{equation}{\rm a}}
\be
 (D_{\bar y}-\lambda D_z)\phi_+=(D_{\bar y}-\lambda D_z)\phi_-,
\ee
\setcounter{equation}{11}
\renewcommand{\theequation}{\arabic{equation}{\rm b}}
\be
(D_{\bar z}+\lambda D_y)\phi_+=(D_{\bar z}+\lambda D_y)\phi_-.
\ee
The action of the algebra $C^1({{\mathfrak U}}, {\mathcal H})$ on
$SL(n, {\mathbb C})$-valued {}functions $\psi_\pm$ and on gauge
potentials $\{A_\mu\}$ is given by the formulae
\setcounter{equation}{12}
\renewcommand{\theequation}{\arabic{equation}}
\be
\delta_\varphi\psi_+=-\phi_+\psi_+,\qquad
\delta_\varphi\psi_-=-\phi_-\psi_- ,
\ee
\setcounter{equation}{13}
\renewcommand{\theequation}{\arabic{equation}{\rm a}}
\be
\delta_\varphi A_{\bar y}-\lambda\delta_\varphi A_z=
D_{\bar y}\phi_+-\lambda D_z\phi_+ =D_{\bar y}\phi_--\lambda
D_z\phi_- ,
\ee
\setcounter{equation}{13}
\renewcommand{\theequation}{\arabic{equation}{\rm b}}
\be
\delta_\varphi A_{\bar z}+\lambda\delta_\varphi A_y=
D_{\bar z}\phi_++\lambda D_y\phi_+ =D_{\bar z}\phi_-+\lambda
D_y\phi_- .
\ee
It follows from (14) that
\setcounter{equation}{14}
\renewcommand{\theequation}{\arabic{equation}}
\be
\ba{l}
\ds \delta_\varphi A_y=\oint_{S^1}\frac{d\lambda}{2\pi i\lambda^2}
(D_{\bar z}\phi_+ +\lambda D_y\phi_+ ) ,\qquad
\delta_\varphi A_z=-\oint_{S^1}\frac{d\lambda}{2\pi i\lambda^2}
(D_{\bar y}\phi_+ -\lambda D_z\phi_+ ) ,
\\[6mm]
\ds \delta_\varphi A_{\bar y}=\oint_{S^1}\frac{d\lambda}{2\pi i\lambda}
(D_{\bar y}\phi_+ -\lambda D_z\phi_+ ) ,\qquad
\delta_\varphi A_{\bar z}=\oint_{S^1}\frac{d\lambda}{2\pi i\lambda}
(D_{\bar z}\phi_+ +\lambda D_y\phi_+ ),
\ea\!\!\!
\ee
where $S^1=\{\lambda\in CP^1 : |\lambda |=1\}$. Thus, we have
described the action of $C^1({{\mathfrak U}}, {\mathcal H})$ on the
space of solutions of SDYM equations.

\medskip

\noindent
{\bf Example 1.} For $\varphi =0$ we have $\phi =0$. Choose
$\phi_+=\phi_-=\vartheta(x)$, $x\in{\mathbb R}^4$, then formulae~(15)
give us manifest gauge symmetries (2).

\medskip

\noindent
{\bf Example 2.} If we choose $\varphi =\varphi (\lambda )$ (i.e.
$\p_\mu\varphi (x,\lambda )=0$), then obtain the action of the
algebra $su(n)\otimes C[\lambda ,\lambda^{-1}]$ on the space of
solutions of SDYM equations~\cite{ivanova:P-D}.

\section{Inf\/initesimal dif\/feomorphism-type symmetries}

{\bf 5.1. The algebra $C^0({\mathfrak U}, {\mathcal V})$.} Let us
consider a complexif\/ied tangent bundle $T^{{\mathbb C}}({{\mathcal Z}
}) = T^{(1,0)}({{\mathcal Z} }) \oplus T^{(0,1)}({{\mathcal Z} })$ of
the twistor space ${\mathcal Z} $ and a sheaf ${\mathcal V}$ of germs
of holomorphic sections of the bundle $T^{(1,0)}({{\mathcal Z} })$.
The set of all sections of the sheaf ${\mathcal V}$ over an open set
${{\mathcal U} }\subset {{\mathcal Z} }$ is denoted by $\Gamma
({{\mathcal U} }, {{\mathcal V} })$.  If we take sections of
${\mathcal V} $ over each of the open sets $U_+$ and $U_-$ from the
cover ${\mathfrak U} $, then the resulting collection of holomorphic
vector f\/ields is called a 0-cochain over ${\mathcal Z} $, subordinate
to the cover ${\mathfrak U} $. Thus, a 0-cochain $\{\eta_+, \eta_-\}$
is an element of the space
\[
 C^0({{\mathfrak U} }, {{\mathcal V}
}):= \Gamma (U_+, {{\mathcal V} }) \oplus
\Gamma (U_-, {{\mathcal V} }).
\]
The space of 1-cochains is def\/ined as follows: $C^1({{\mathfrak U} },
{{\mathcal V} }):= \Gamma (U, {{\mathcal V} }), $ where $U=U_+\cap
U_-$. Thus, elements of $C^1({{\mathfrak U} }, {{\mathcal V} })$ are
holomorphic vector f\/ields $\eta_{+-}$ def\/ined on $U$.

\medskip

\noindent
{\bf 5.2. Action of $C^0({{\mathfrak U} }, {{\mathcal V} })$ on
transition matrices.} The vector space $C^0({{\mathfrak U}},
{\mathcal V})$ can be described as the Lie algebra of holomorphic
vector f\/ields with pointwise commutator, def\/ined on $U_+$ and $U_-$.
For any $\eta =\{\eta_+ ,\eta_-\}\in C^0({{\mathfrak U}}, {\mathcal
V})$ we def\/ine two actions of $C^0({{\mathfrak U}}, {\mathcal V})$ on
the transition matrix ${\mathcal F}$
\be
\delta^\pm_{\eta}{\mathcal F} = \eta_\pm ({\mathcal F}),
\ee
i.e. as a derivative of ${\mathcal F}$ along the vector f\/ields
$\eta_\pm \in C^0({{\mathfrak U}}, {\mathcal V})$.

One can also consider a combination of these actions:
\[
\delta_\eta{\mathcal F} =\delta_\eta^-{\mathcal F}
-\delta_\eta^+{\mathcal F}.
\]
It is easy to see that the algebra $C^0({{\mathfrak U}}, {\mathcal
V})$ acts on the algebra $C^1({{\mathfrak U}}, {\mathcal H})$ by
derivations, and we can consider a semidirect sum $C^0({{\mathfrak
U}}, {\mathcal V})\dotplus C^1({{\mathfrak U}}, {\mathcal H})$ of
these algebras.

\medskip

\noindent
{\bf 5.3. Action of $C^0({{\mathfrak U}}, {\mathcal V})$ on self-dual
connections.} Let us introduce the $sl(n,C)$-valued smooth functions
$\theta^\pm$ on $U$
\[
\theta^\pm :=\psi_-(\delta_\eta^\pm {\mathcal F})\psi_{+}^{-1},
\]
which are holomorphic in $\lambda\in C_{\alpha}$:
\[
\theta^\pm =\sum_{n=-\infty}^{\infty}\lambda^n\theta_n^\pm (x)
=\theta_-^\pm -\theta_+^\pm ,
\]
where
\[
\ds \theta_+^\pm :=\tilde\theta_{0}^\pm (x)-\!
\sum\limits_{n=1}^{\infty}\lambda^n\theta_n^\pm (x),\quad \!\!
\theta_-^\pm :=\hat\theta_{0}^\pm (x)+\!\!
\sum\limits_{n=-\infty}^{-1}\lambda^n\theta_n^\pm (x),\quad\!\!
\ds \hat\theta_{0}^\pm (x)-\tilde\theta_{0}^\pm (x)=\theta_{0}^\pm (x).
\]
 Thus, the functions $\theta_\pm^\pm (x,\lambda )\in sl(n,C)$ are
holomorphic in $\lambda^{\pm 1}\in C_\pm\subset {\mathbb C}\!P^1$.

For $\theta_-^\pm $ and $\theta_+^\pm $ we have
\setcounter{equation}{16}
\renewcommand{\theequation}{\arabic{equation}{\rm a}}
\be
(D_{\bar y}-\lambda D_z)\theta_+^\pm =(D_{\bar y}-\lambda
D_z)\theta_-^\pm ,
\ee
\setcounter{equation}{16}
\renewcommand{\theequation}{\arabic{equation}{\rm b}}
\be
(D_{\bar z}+\lambda D_y)\theta_+^\pm =(D_{\bar z}+\lambda
D_y)\theta_-^\pm .
\ee
The action of $C^0({{\mathfrak U}}, {\mathcal V})$ on
matrix-valued functions $\psi_\pm \in SL(n,C)$ and on gauge
potentials $\{A_\mu\}$ is given by the formulae
\setcounter{equation}{17}
\renewcommand{\theequation}{\arabic{equation}}
\be
\delta_\eta^\pm \psi_+:=-\theta_+^\pm \psi_+,\qquad
\delta_\eta^\pm \psi_-:=-\theta_-^\pm \psi_- ,
\ee
\setcounter{equation}{18}
\renewcommand{\theequation}{\arabic{equation}{\rm a}}
\be
\delta_\eta^\pm  A_{\bar y}-\lambda\delta_\eta^\pm  A_z:=
D_{\bar y}\theta_+^\pm -\lambda D_z\theta_+^\pm =D_{\bar
y}\theta_-^\pm -\lambda D_z\theta_-^\pm ,
\ee
\setcounter{equation}{18}
\renewcommand{\theequation}{\arabic{equation}{\rm b}}
\be
\delta_\eta^\pm  A_{\bar z}+\lambda\delta_\eta^\pm  A_y:=
D_{\bar z}\theta_+^\pm +\lambda D_y\theta_+^\pm =D_{\bar
z}\theta_-^\pm +\lambda D_y\theta_-^\pm .
\ee
It follows from (19) that
\setcounter{equation}{19}
\renewcommand{\theequation}{\arabic{equation}}
\be
\ba{l}
\ds \delta_\eta^\pm  A_y=\oint_{S^1}\frac{d\lambda}{2\pi i\lambda^2}
(D_{\bar z}\theta_+^\pm +\lambda D_y\theta_+^\pm ) ,\qquad
\delta_\eta^\pm  A_z=-\oint_{S^1}\frac{d\lambda}{2\pi i\lambda^2}
(D_{\bar y}\theta_+^\pm -\lambda D_z\theta_+^\pm ) , \\[6mm]
\ds \delta_\eta^\pm  A_{\bar y}=\oint_{S^1}\frac{d\lambda}{2\pi i\lambda}
(D_{\bar y}\theta_+^\pm -\lambda D_z\theta_+^\pm ) ,\qquad
\delta_\eta^\pm  A_{\bar z}=\oint_{S^1}\frac{d\lambda}{2\pi i\lambda}
(D_{\bar z}\theta_+^\pm +\lambda D_y\theta_+^\pm ),
\ea\!\!
\ee
where $S^1=\{\lambda\in {\mathbb C}\!P^1 : |\lambda |=1\}$.

\medskip

\noindent
{\bf Example 3.} Let us consider the holomorphic vector f\/ields $\eta
=\lambda^{-n}\tilde N$ on $U\subset{\mathcal Z}$, $n=0, \pm 1, \pm
2,\ldots$,  where $\tilde N$ are vector f\/ields on ${\mathcal Z}$
realizing the action of $so(5,1)$ on ${\mathcal Z}$, which preserves
the holomorphicity of the bundle $\tilde E \rightarrow {\mathcal Z}$.
Such lift $N\rightarrow\tilde N$ of vector f\/ields from~${\mathbb
R}^4$ to ${\mathcal Z}$ was described in~\cite{ivanova:PL}. As it has been
shown in~\cite{ivanova:ita}, symmetries (20) for $\eta =\lambda^{-n}\tilde N$
with $n\ge 0$ are in one-to-one correspondence with the symmetries
from~\cite{ivanova:PP}.

\section{Conclusion}

To sum up, using the one-to-one correspondence between the classes of
holomorphically equivalent transition matrices ${\mathcal F}$ and the
gauge equivalent classes of self-dual connections, to any
inf\/initesimal transformations (11) and (16) of transition matrices we
have associated the inf\/initesimal transformations (15) and (20) of
solutions $\{A_\mu\}$ of the SDYM equations. There are no other
inf\/initesimal automorphisms of the bundle $\tilde E$ over ${\mathcal
Z}$ besides those generated by the algebras $C^0({{\mathfrak U}},
{\mathcal V})$ and $C^1({{\mathfrak U}}, {\mathcal H})$.  Thus, the
inf\/inite-dimensional algebra of all inf\/initesimal transformations of
solutions of the SDYM equations has the form $C^0({{\mathfrak U}},
{\mathcal V})\dotplus C^1({\mathfrak U}, {\mathcal H})$.

Notice that all the results of this paper may be generalized to the
case of the SDYM equations in $4n$-dimensional spaces considered e.g.
in \cite{ivanova:W-P}. It would be interesting to generalize our results to
other SDYM-type equations in dimension greater than four (see
e.g.~\cite{ivanova:CFDN}), various modif\/ications of which arise in string and
membrane theories (see e.g.~\cite{ivanova:Har} and references therein).

\subsection*{Acknowledgements}

The author is grateful to Ina Kersten, Sylvie Paycha and Sheung Tsun
Tsou for encouraging and helpful discussions. This work is supported
in part by the grant No. 98-01-00173.

 \label{ivanova-lp}
\end{document}